\documentstyle[preprint,aps,epsfig]{revtex}

\def\be{\begin{equation}}
\def\th0{\theta_0}

\def\bea{\begin{eqnarray}}
\def\ee{\end{equation}}
\def\eea{\end{eqnarray}}

\begin{document}
\title{Gas of self-avoiding loops on the brickwork lattice}

\author{F. Eghbal and D. P. Foster \\
{\em Laboratoire de Physique Th\'eorique et Mod\'elisation, \\
Universit\'e de Cergy-Pontoise,
2 ave A. Chauvin, 95302 Cergy-Pontoise, France} \\
and \\
H. Orland  \\
{\em Service de Physique Th\'eorique, \\
C.E. Saclay, Orme des
Merisiers, 91911 Gif-sur-Yvette, France}
 }
%

\date{June 14, 1997}
\maketitle

\begin{abstract}
An exact calculation of the phase diagram for a loop gas model on the
brickwork lattice is presented. The model includes a bending energy.
In the dense limit, where all the lattice sites are
occupied, a phase transition occuring at an asymmetric Lifshitz
tricritical point
is observed as the temperature
associated with the bending
energy is varied.  Various critical exponents are calculated. At lower
densities, two lines of transitions (in the Ising universality class)
are observed, terminated by a tricritical point, where there is a
change in the modulation of the correlation function. To each
tricritical point an associated disorder
line is found.

\end{abstract}
\vskip 2cm
\noindent\mbox{Submitted for publication to:} \hfill 
\mbox{Saclay, SPhT/97-xxx}\\ \noindent \mbox{``Journal of Physics A''}\\ 
\noindent \mbox{PACS: 05.50.+q, 64.60.Kw}

\newpage

\section{Introduction}

Models of closed loops on lattices in two dimensions have attracted
considerable attention \cite{Nienhuis}. 
In a theoretical context they arise naturally
as high-temperature expansions of spin models and they are closely related
to integrable systems such as vertex models \cite{Baxter}. 
Loops on a lattice may
also be regarded as simple models for (short) ring polymers in
solution \cite{deGennes,desCloizeaux}. The
segments of the loop are then regarded as monomers, or small clusters
of monomers. While realistic systems are three-dimensional, the
two-dimensional case provides rich critical behaviour, and
it may be hoped that some features hold in higher dimensions. Solid-on-solid
models used in the study of roughening transitions in
three-dimensional growth may also be mapped onto various
two-dimensional loop-gas models \cite{Nienhuis}.

In this paper a model of loops is studied, consisting of self-avoiding
rings with a
bending energy. We are mainly interested in the effects of varying the
density of monomers on the lattice and the temperature.
The model is defined as follows. Each bond of the lattice is either
occupied by a monomer or empty. Each monomer placed on the lattice connects
to two others such that the only allowed configurations
consist of closed self-avoiding loops. An energy penalty $\epsilon$ is
associated with each turn. The density of lattice bonds occupied is
allowed to change, and the model is studied in the grand canonical
ensemble by introducing a fugacity $K$ for each monomer.

On the square lattice, and in the limit that the lattice is maximally
occupied (all the sites visited) this model corresponds to the
so-called F~Model \cite{Wu},
which exhibits an infinite order phase transition as the bending
energy, or equivalently temperature, is varied. At low temperatures
corners are expelled from the bulk, while at high temperatures there is
a proliferation of corners.

A qualitatively similar transition is seen in a model of a single
Hamiltonian walk on the square lattice with a bending energy \cite{Flory}.
The Hamiltonian walk may be
thought of as the limit of an interacting self-avoiding walk where the
attractive nearest-neighbour monomer-monomer interactions are strong
enough to exclude any lattice vacancies. In the other limit, the
non-interacting self-avoiding walk, it is known that the bending
energy is irrelevant \cite{deGennes,desCloizeaux}; 
it changes the effective size of a monomer without
changing the critical behaviour. At low enough densities we expect
that for the loop-gas model the bending energy will also be irrelevant
in this sense.

One therefore expects that between zero density (bending interaction
irrelevant) and density one (bending interaction relevant) there
should be a ``critical'' density corresponding to a change of
behaviour.

In this paper we study the loop gas model on the brickwork
lattice. The brickwork lattice corresponds to a square lattice with
half the horizontal bonds removed, giving rise to the ``brick wall''
motif, see Figure~1. The brickwork lattice is topologically identical
to the hexagonal lattice. 


At density one we find a qualitatively similar phase transition
to that on the standard square lattice in that there is a low
temperature phase in which the corners are expelled from the bulk and
a high temperature phase where there is a finite density of corners in
the bulk. The details of the transition are however very different:
the transition occurs at a tricritical Lifschitz point and the 
high-temperature phase is modulated and critical. Such a 
phase transition is
reminiscent of the Pokrovsky-Talapov transition
\cite{Pokrovski_Talapov}. 
In this limit our
model is
equivalent to a modified KDP model on the hexagonal lattice \cite{wu,Blote}.
The existence of a modulated phase at $\rho=1$
implies the existence of lines of disorder in the phase diagram
\cite{Stephenson},
along which the correlation functions change from modulated at high
densities to monotonic at low densities.

As the density is reduced the tricritical point is extended into a
line of critical points in the Ising universality class which
terminates at zero temperature at a critical density of about
0.8. For $K<1$, or at low density, another line of critical
points is observed, again in the Ising universality class; our model
in this region is essentially the Ising model on the hexagonal
lattice with one of the three couplings different from the other two.

The remainder of this paper is organised as follows. In
Section~\ref{II}, the
grand canonical partition function (and other relevant quantities) is 
calculated by expressing
it first in terms of Grassman integrals, which are then exactly
computed in the thermodynamic limit. In Section~\ref{III}  the
$K\to \infty$ ($\rho\to 1$) limit is discussed, along with the
nature of the low and high temperature phases. In Section~\ref{IV} the phase
boundaries and lines of disorder are found and the different aspects
of the phase diagram discussed. Section~\ref{V} is devoted to final
discussions and conclusions.

\section {The model}
\label{II}
We consider a two dimensional gas of loops on a brickwork lattice (BW). The
loops are self-avoiding, and we assign a fugacity $K$ to each occupied
link.
This BW lattice
can be visualized as a honeycomb (HC) lattice (see
Fig.~1), and we associate a weight of $e^{- \beta \varepsilon}$
(where $\beta = 1/T$ is the inverse temperature)
to each
corner of a loop 
or equivalently $\lambda=e^{-2 \beta \varepsilon}$ to horizontal
links. 
This model is a straightforward extension of the F-model to
the BW lattice. The aim of this paper is to calculate the phase
diagram and properties of this system as a function of the site
density $\rho$ (or equivalently bond fugacity $K$) and temperature.
The grand canonical partition function $Z$ of the system is given by:
\bea
\label{grand_canon}
Z&=&\sum_{N=0}^\Omega K^N Z_N \\
&=& e^{-\beta \Omega f(K,T)}
\eea
where $\Omega= L^2$ is the total number of sites of the lattice of
linear dimension $L$, and $Z_N$ is the $N-$site partition function.
As usual, the canonical partition function at site density
$\rho = {N / \Omega}$
can be obtained through
(\ref{grand_canon}) by:
\bea
\label{canon}
Z_N&=&\oint {dK \over K} e^{-\Omega [\rho \log K -\beta f(K,T)]} \\
&=&e^{-\beta \Omega f_c(\rho,T)}
\eea
In the thermodynamic limit, we have the usual relation
\be
\rho = K {\partial \over \partial K} \log Z
\ee

The loop gas can be identified with the graphs of the high-temperature
expansion of an Ising model on a BW lattice, with a weight $K$ per vertical
bond and $\lambda^2 K$ per horizontal bond. This identification holds
only provided that $K \le 1$. However, the solution follows for any
value of $K$.

Using the results of Houtappel \cite{Houtappel,Syozi}, we have, for
any $K$:
\be
\label{grand_pot1}
f(K,T)={1\over {16 \pi^2}} \int_0^{\pi} dk_x \int_0^{\pi} dk_y \log
(a_1 K^4+a_2 K^2 +a_3)
\ee
where
\bea
a1&=& 1+4 \lambda^4 \cos^2 k_y -4 \lambda^2 \cos k_x \cos k_y  \\
a2&=& 2 - 4 \cos^2 k_y + 4 \lambda^2 \cos k_x \cos k_y \\
a3&=& 1 \\
\eea

This result as well as the correlation functions can be easily
obtained through the use of Grassman variables.
Following Samuel \cite{Samuel,Dotsenko},
the partition function $Z$ can be represented as a Grassman integral
\be
\label{grass}
Z=\int {\cal D} \psi {\cal D} \varphi e^{-A}
\ee
where
\bea
\label{grass1}
A=&-&\sum_{m,n} \left( \psi_3^{m,n} \psi_1^{m,n}+ \psi_4^{m,n}
\psi_2^{m,n}+ 
\varphi_3^{m,n} \varphi_1^{m,n}+ \varphi_4^{m,n} \varphi_2^{m,n}\right. \\ \nonumber
&&+\lambda (\psi_1^{m,n} \psi_2^{m,n}+ \psi_3^{m,n} \psi_4^{m,n}+
\psi_2^{m,n} 
\psi_3^{m,n}+ \psi_1^{m,n} \psi_4^{m,n} \\ \nonumber
&&\hspace{.7cm} \varphi_1^{m,n} \varphi_2^{m,n}+ \varphi_3^{m,n} 
\varphi_4^{m,n}+ \varphi_2^{m,n} \varphi_3^{m,n}+ \varphi_1^{m,n} 
\varphi_4^{m,n}) \\ \nonumber
&&+ K(\lambda^2
\left. \varphi_3^{m,n}\psi_1^{m,n}+\psi_4^{m,n+1}\varphi_2^{m,n}+
\varphi_4^{m,n}\psi_2^{m-1,n}) \right)
\eea
and $\psi_1^{m,n}, \psi_2^{m,n}, \psi_3^{m,n}, \psi_4^{m,n}$
are Fermionic fields attached to each lattice site $(m,n)$.
The Fermion integral can be performed and the grand potential
\ref{grand_pot1} can be recovered. In addition, the 
generic correlation functions read:
\be
G(m,n)= (-1)^m {1\over {4 \pi^2}} \int_0^{\pi} dk_x \int_0^{\pi} dk_y { e^{i
(m k_x + n k_y)} \over a_1 K^4+a_2 K^2 +a_3 }
\ee
The actual correlation functions contain regular multiplication
factors which do not modify the long distance behaviour.

Integration over $k_x$ can be performed and gives:
\bea
\label{grand_pot2}
f(K,T)&=&{1\over{8 \pi}} \int_0^{\pi} dk_y \log
\frac{a+\sqrt{a^2-b^2}}{2} \\
\label{corr2}
G(m,n)&=& (-1)^m \int_0^{\pi} {dk_y \over 2 \pi} {e^{i n k_y} \over
\sqrt{a^2-b^2}}
\left(
{ \sqrt{a^2 - b^2} -a \over b} \right) ^m
\eea
where

\bea
a&=& (1+4 \lambda^4 \cos^2 k_y) K^4 + (2-4 \cos^2 k_y) K^2 +1 \\
b&=&-4 \lambda^2 \cos k_y K^2 (K^2-1)
\eea

The canonical free energy (at bond density $\rho= N/ \Omega$) is given
by:
\be
\label{free}
f_c(\rho,T) = \rho \log K - f
\ee
where $K$ is determined 
as a function of $\rho$ through:

\be
\label{density}
{\rho \over K} = {1\over{8 \pi}} \int_0^\pi
dk_y {1\over {g(K,T,k_y)}} \frac{\partial g(K,T,k_y)}{\partial K}
\ee
and
\be
g=\frac{a+\sqrt{a^2-b^2}}{2}
\ee

In the following, we will work in the grand canonical ensemble, and
transpose the results to the canonical ensemble when necessary. We
first consider the fully-packed lattice ($\rho=1$) and then discuss
the dilute case.

\section{The fully-packed lattice}
\label{III}
By analogy with polymer theory, it is interesting to consider the case
where all lattice sites are visited once and only once by the
loops. This is the non-connected version of Hamiltonian path model, with
a penalty factor $\lambda$ per corner. From Eq.~(\ref{density}), we see
that $\rho=1$ for $K= \infty$. We are thus led to study
Eq.~(\ref{free}) in the limit when $K \to \infty$. One obtains:

\be
\label{rho_1}
f_c=-{1\over{4 \pi \beta}} \int_0^\pi dk_x \int_0^\pi dk_y
\log(1+4\lambda^4 \cos^2 k_y - 4\lambda^2 \cos k_x \cos k_y)
\ee

As usual we identify the critical points of the system from the zeros
of the argument of the log in the above equation. It may be seen that
no zeros exist for $\lambda<1/\sqrt {2}$ and that for $\lambda\ge 1/\sqrt{2}$
zeros
exist at:
\bea
\label{k_crit}
k_x&=&0 \\ \nonumber
\cos k_y &=& {1 \over 2 \lambda ^2}
\eea

This implies that the whole region $\lambda\ge 1/\sqrt{2}$ is critical, with
a temperature dependent critical wavevector. We therefore identify
$\lambda=1/\sqrt{2}$ with a tricritical Lifshitz point, and the region
$\lambda \ge 1/\sqrt{2}$ as a Lifshitz line of critical points. 
Using the definition $\lambda = \exp{- \varepsilon / T}$, 
the
corresponding temperature for the tricritical point is
$T_c=2\varepsilon/\log 2$.

The integration over $k_x$ in Eq.~(\ref{rho_1}) may be carried out
explicitly, giving:
\be
\label{rho=1}
f_c=-{1\over{4 \pi \beta}} \int_0^\pi dk_y \log(\frac{1+4 \lambda^4
\cos^2 k_y + |1-4\lambda^4 \cos^2 k_y|}{2})
\ee
or
\bea
\label{full}
f_c&=&-{1\over2} \int_0^{\arccos (1/2 \lambda^2)} dk_y \log(4 \lambda^4 \cos^2 k_y) \hspace{.8cm} {\rm for} \hspace{.8cm} T\geq T_c \\
&\equiv& 0 \hspace{3.8cm} {\rm for} \hspace{3.8cm} T \leq T_c \\
\eea
This form is similar to the models studied by
\cite{Pokrovski_Talapov,wu} (see Fig.~2).

It is natural to define the average number density
of corners $n_c$ as the order
parameter for this transition. Indeed, we find (see Fig.~3):
\bea
\label{order}
n_c&=& {2\over\pi} \arccos ({1\over{2 \lambda^2}}) \hspace{1.cm}{\rm for}
\hspace{1.cm} T\geq T_c \\
&\equiv& 0 \hspace{3.3cm}{\rm for}\hspace{1.1cm} T \leq T_c
\eea

The critical behaviour of the order parameter is given by $n_c
 \sim
\delta T ^{1/2}$  as $T \to T_c$
so that the critical exponent $\beta$ is equal to
1/2.
The low-temperature phase is completely frozen, consisting of straight
vertical lines, with all the corners
rejected to the outer boundary. The high-temperature phase is
modulated in the $y$-direction with a wave vector given by
Eq.~(\ref{k_crit})

The same critical behaviour is also seen in the zero-temperature
phase diagram of the frustrated Ising model on the triangular lattice with
appropriately chosen coupling constants \cite{wu,Blote}.
This
model may be mapped onto a
tiling consisting of three types of lozenge \cite{Blote}.
One lozenge has a
lower energy than the other two.
At zero temperature, the tiling must be perfect. One rapidly realises
that the only way of introducing a lozenge of higher energy is to
introduce an infinite line of them.
Identifying the side of a lozenge with the bisector of an occupied
bond on the dual hexagonal lattice, our loop model may be seen as
equivalent to this lozenge tiling (see Fig.~4). 
At non-zero temperature (as defined
in our model) a defect line may be seen as a restricted SOS
interface crossing the lattice. The energy needed to create one such
line is $E=2\varepsilon L$
and the entropy is $S=L\log 2$. When the
associated free energy, $F_1=L(2\varepsilon-T\log 2)$, becomes negative,
defect lines (and hence corners) proliferate. This defines the
critical temperature as $T_c=2\varepsilon/\log 2$, consistent with the
tricritical temperature found analytically above.

In the high-temperature phase, where these lines proliferate, we give
a simple physical argument for the observed modulation in the
correlation functions. The free energy $F_1$ is simply the chemical
potential for creating one such line. When a finite density of lines
is present, the reduction of entropy must be taken into
account\cite{helfrich}, yielding an effective repulsion between
them. The total free energy for $N$ lines, per occupied bond, is:
\be
F_{N}=-F_1 N+c \sum_{i=1}^{N} {T \over d_{i}^2},
\ee
where $d_i$ is the distance between lines $i$ and $i+1$. Minimising
$F_N$ with respect to the $d_i$, subject to the constraint $\sum_i
d_i=L$, gives all the $d_i$ equal and given by:
\be
d_i\sim{1 \over \sqrt{T\log 2-2\varepsilon}}
\ee
explaining the form of the temperature dependence of the modulation wavevector,
Eq.~(\ref{k_crit})

Close to the tricritical Lifshitz point, the free energy scales as:
\be
f \sim \frac{4}{\pi T_c^3} (\delta T)^{3/2}
\ee
from which we obtain the specific-heat critical exponent $\alpha=1/2$.

Along the Lifshitz line, the critical
behaviour of the
correlation functions can be analysed; a generic correlation
function is given by:
\be
G(\vec r) = {1 \over {4 \pi^2}} \int_0^\pi \int_0^\pi d^2 k
\frac{e^{-i {\vec k}.{\vec r} }}{1+4\lambda^4 \cos^2 k_y - 4\lambda^2 \cos k_x \cos k_y}
\ee
up to a regular multiplication factor.

Away from the tricritical point, i.e. $\lambda > 1/\sqrt{2}$,
we may develop  around the
critical wave vector $(0,k)$ defined in Eq.~(\ref{k_crit}) Setting
$k_x=q_x$ and $k_y=k+q_y$,
this becomes:
\be
G(x,y) \sim e^{-i k y} \int d q_x \int d q_y
\frac{e^{-i {\vec q}.{\vec r} }}{q_x^2+(4 \lambda^4 -1) q_y^2}
\ee
The exponential prefactor gives the expected spatial modulation in the
y-direction, and the correlation function has a logarithmic behaviour
at large distances.

Around the tricritical point, the critical wave vector vanishes as well as
the coefficient of the $q_y^2$ term,
and the expansion must be carried to the next order:
\be
G(x,y) \sim \int d q_x \int d q_y
\frac{e^{-i {\vec q}.{\vec r} }}{\mu^2+q_x^2 - \mu q_y^2+ q_y^4/4}
\ee
where $\mu=2 \lambda^2 -1$. Therefore,
the correlation functions have anisotropic scaling, with critical
exponents $\nu_x = 1$ and $\nu_y = 1/2$.

\section{The dilute lattice}
\label{IV}
We now move to the dilute case $\rho < 1$.

\subsection{Critical lines}
As the density $\rho$ or the fugacity $K$ is lowered,
the tricritical Lifshitz point extends into a critical line. This line
can be obtained from the zeros of the logarithm of
Eq.~(\ref{grand_pot1})
\bea
k_x &=& 0 \nonumber \\
k_y &=& 0 \nonumber \\
K &=& {1 \over \sqrt {1 - 2 \lambda^2}}
\eea
This critical line exists for $K \ge 1$.
In the $\rho-T$ plane, its equation close to
the fully-packed case $\rho=1$ is given by:
\be
\rho \simeq 1 + 2\varepsilon {\delta T \over T_c^2}
\ee
where $T_c = 2\varepsilon/\log 2$ is the Lifshitz tricritical temperature.

For $K \le 1$, there exists another critical line given by:
\bea
k_x &=& 0 \nonumber \\
k_y &=& \pi \nonumber \\
K &=& {1 \over \sqrt {1 + 2 \lambda^2}}
\eea
In this region ($K <1$), the fugacity can be identified with the $\tanh
\beta J_1$ of a regular anisotropic Ising model on a
honeycomb lattice. Similarly, the second coupling is given by $
\lambda^2 K = \tanh \beta J_2$.

It can be easily seen that both lines correspond to the $2d$ Ising
universality class : $\nu=1$. Note that since the problem is
formulated as a loop gas, the correlation functions don't correspond
to the spin correlation functions of the Ising model; here we have
$\eta =0$.

The phase diagram in the $K-T$ plane is shown in Fig~5.
Using equation \ref{density}, we find the phase diagram in the
$\rho-T$ plane (see Fig.~6). One can identify a high-density
transition line and a low-density transition line. The high-density
line ends at $\rho_c \simeq 0.8185$ 
and the low-density one at $\rho_c \simeq 0.19$.

The two critical lines merge at $K=1$, where three phases become
critical simultaneously, defining a tricritical point. This is
manifested in the $\rho-T$ plane by a jump in the critical density
$\rho$ at $K=1$. As usual for zero temperature tricritical points,
observables develop essential singularities.

\subsection{Disorder line}

>From Eq.~(\ref{corr2}), it is easily seen that the correlation
functions change from oscillating (in the $x$-direction)
 for $K > 1$ to monotonic for $K <
1$. These two regimes must therefore be separated by a line
where the short distance correlation changes from oscillating to
non-oscillating. This line $K=1$, which passes through the tricritical
point ($K=1, T=0$), is called a disorder line
\cite{Stephenson}. According to the definition of Garel and Maillard 
\cite{garel}, this
is a line of disorder points of the first kind (with zero correlation length).

We have seen that at $\rho=1$, there is a Lifshitz critical point
separating a frozen low-temperature phase from a modulated (in the
$y$-direction)
high-temperature phase.
At lower densities, the correlation functions
are not modulated. This happens separately for each value of $k_x$.
Following Garel and Maillard \cite{garel}, we define the
disorder line as
the line for which the first mode ($k_x=0$) changes behaviour:
\be
K= {1 \over \sqrt{2 \lambda ^2 -1}}
\ee
This line is defined in the high-temperature region $\lambda >
1/\sqrt{2}$ only, and corresponds to a line of disorder points of the
second kind.

\section{Conclusion}
\label{V}
In this article, a loop gas model on a brickwork lattice was
considered. An energetic penalty was included for each corner. At
$\rho=1$ we observed a phase transition from a low-temperature frozen
(corner free) phase to a high-temperature phase modulated in the
$y$-direction. The phase
transition occurs at a tricritical Lifshitz point, where $\nu_x=1,
\nu_y=1/2$. The whole high-temperature phase is critical. These
results are reminiscent of a phase transition of the Pokrovsky-Talapov
type. This behaviour is completely different from the critical behaviour of
the analogously defined model on the square lattice at $\rho=1$ (the
F-Model). This is due to the combination of two effects; the brickwork
lattice automatically imposes self-avoidance without the inclusion of
additional fugacities, and the brickwork lattice is intrinsically
asymmetric.

The phase diagram is given in the $K-T$ (and equivalently the
$\rho-T$) plane.
Two lines of critical points were observed
corresponding to high and low-density phase transitions. The
high-density phase transition is to a phase modulated in the
$x$-direction,
and the low-density phase corresponds to the usual Ising transition.
Both transitions are in
the Ising universality class and meet at $T=0$ at another
tricritical point.

While the model studied is simple, the resulting phase diagram is
suprisingly complex. In the formalism chosen, it is not clear how to
characterise the different finite-density phase transitions in terms
of the loop-model observables.

%
%
%
%
%
%
%
%
%

\begin{figure}
\vskip 1.5cm
\centerline{
\psfig{figure=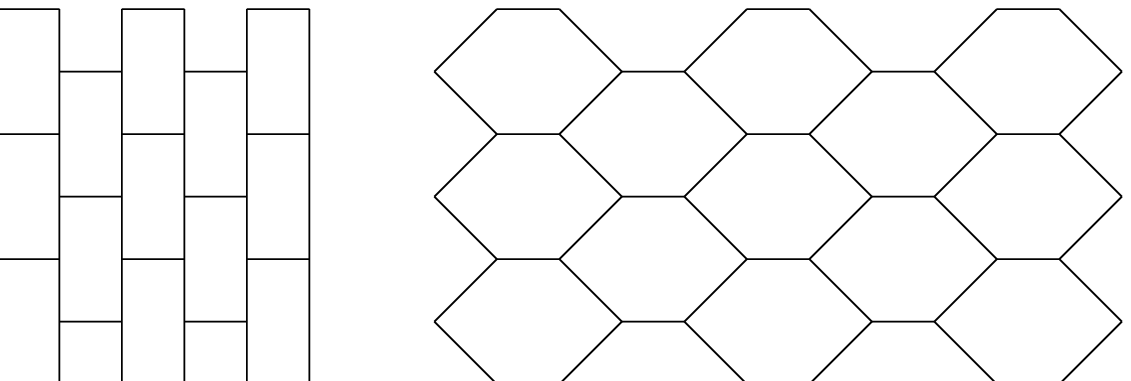,width=6in}}
\caption{The Brickwork and the corresponding Honeycomb lattice.}
\end{figure}

\begin{figure}
\vskip 1cm
\centerline{
\psfig{figure=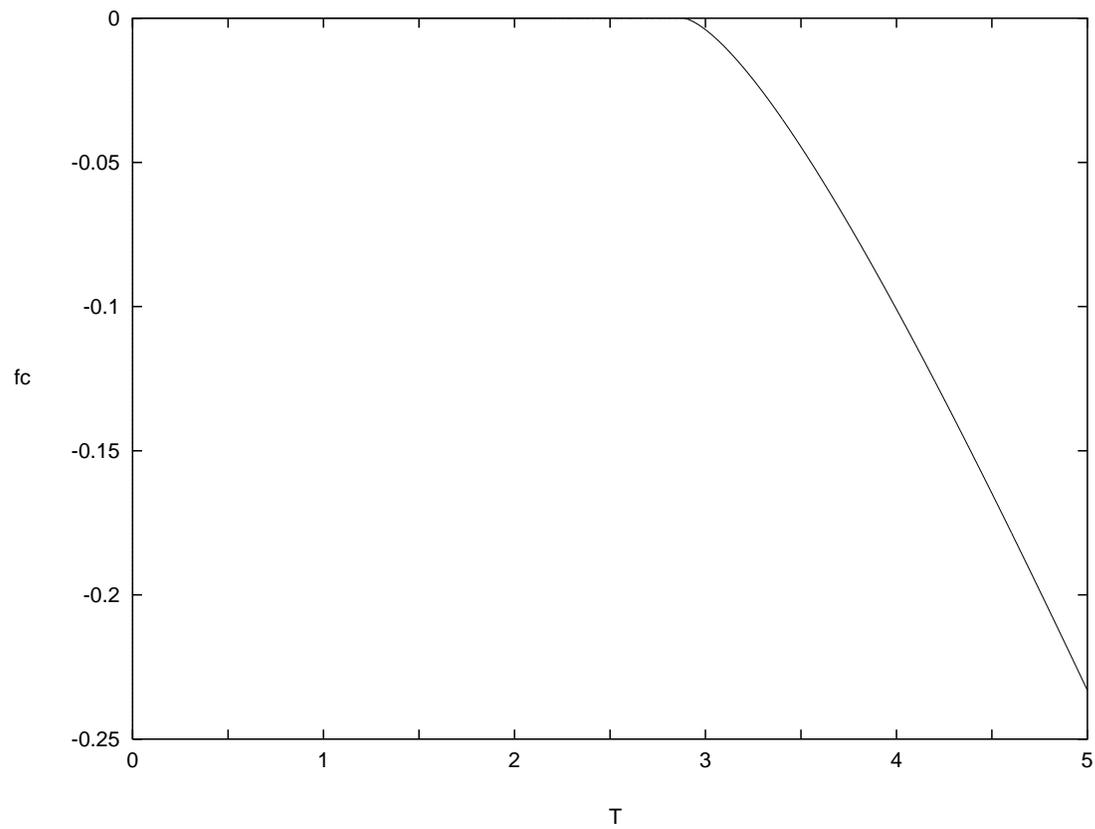,height=4in} }
\vskip 0.5cm
\caption{The free energy as a function of temperature for
the fully packed model.
} 
\end{figure}

\begin{figure}
\vskip 1cm
\centerline{
\psfig{figure=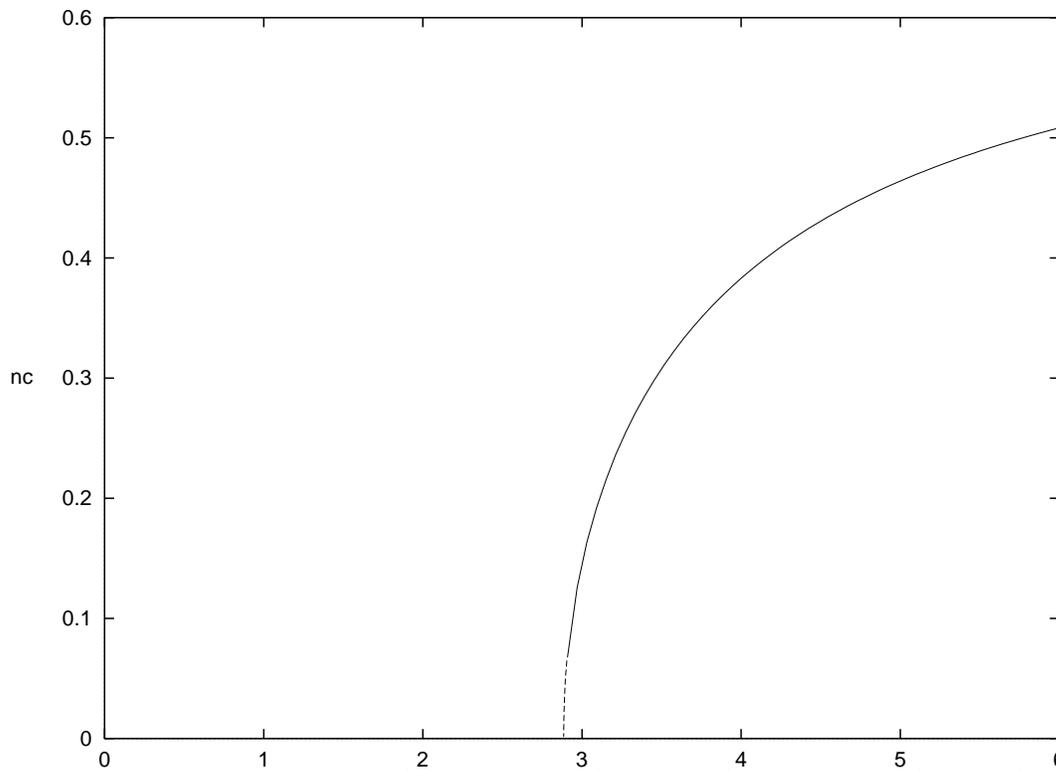,height=4in} }
\caption{
The average number of corners as a function of temperature for 
the fully packed model.
}
\end{figure}
\newpage

\begin{figure}
\vskip 1cm
\centerline{
\psfig{figure=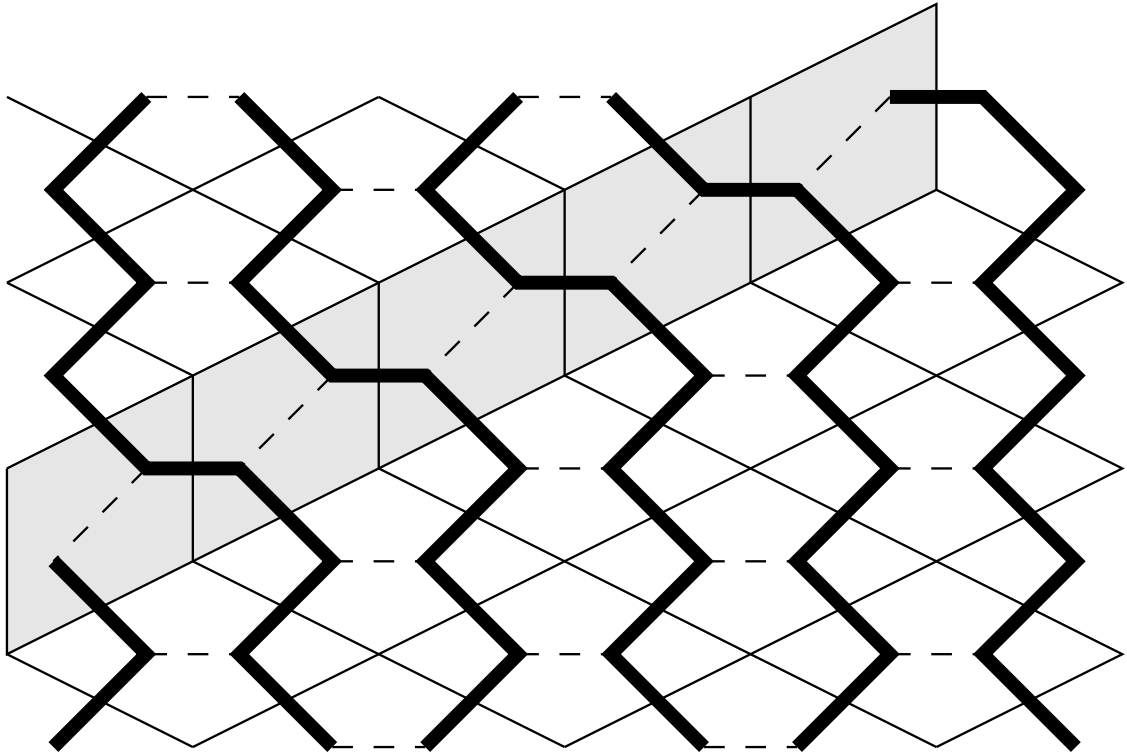,height=4in} }
\caption{The mapping between the loop gas and the lozenge model: the
shaded region shows an infinite line of defects.}
\end{figure}

\begin{figure}
\centerline{
\psfig{figure=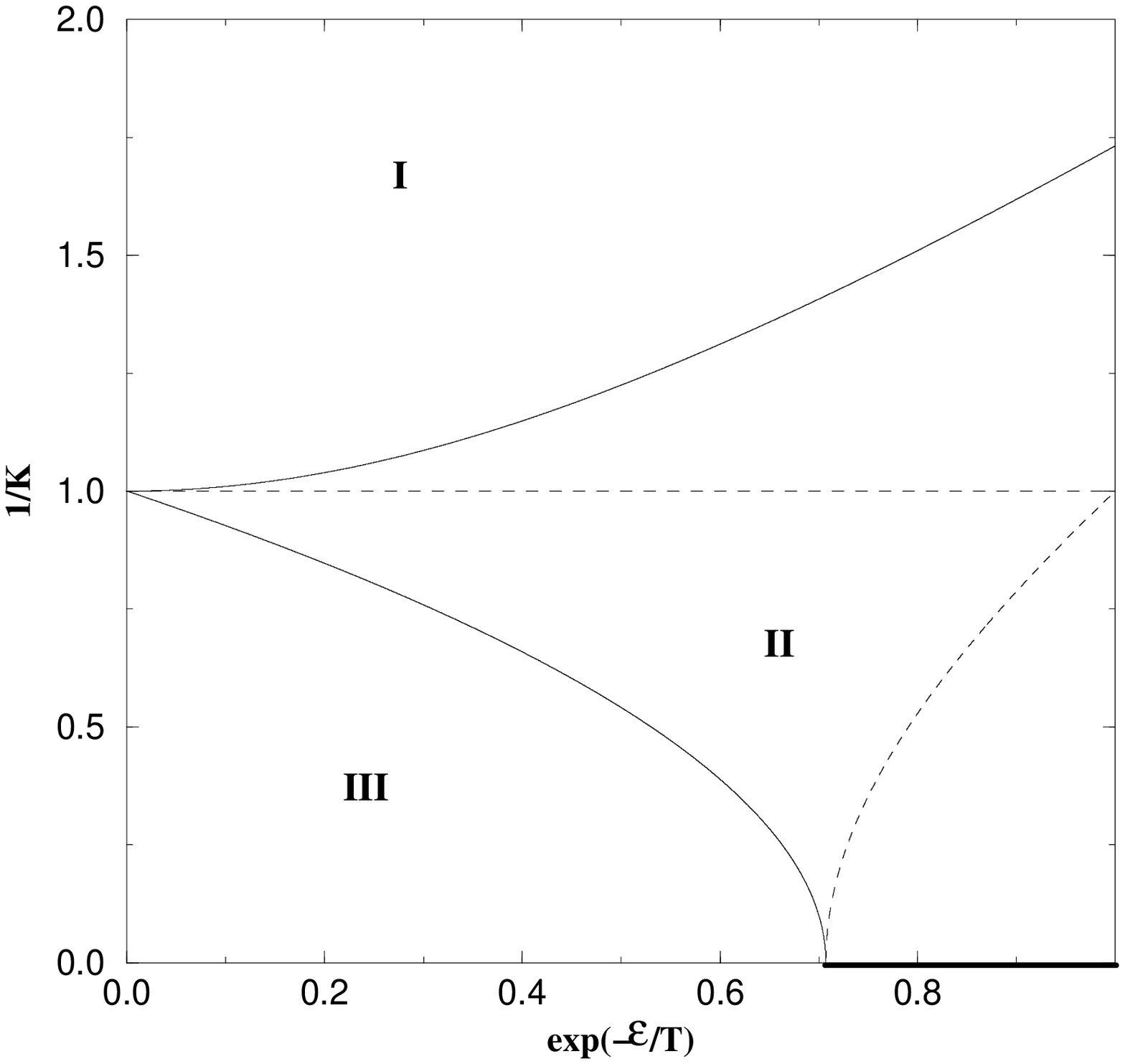,height=5in} }
\caption{Phase diagram of the loop gas in the $K-T$ plane; the solid
lines denote the phase boundaries; the thick line represents the
Lifshitz line; the dashed lines correspond to the disorder lines. The
three phases are defined in the text.}
\end{figure}

\begin{figure}
\centerline{
\psfig{figure=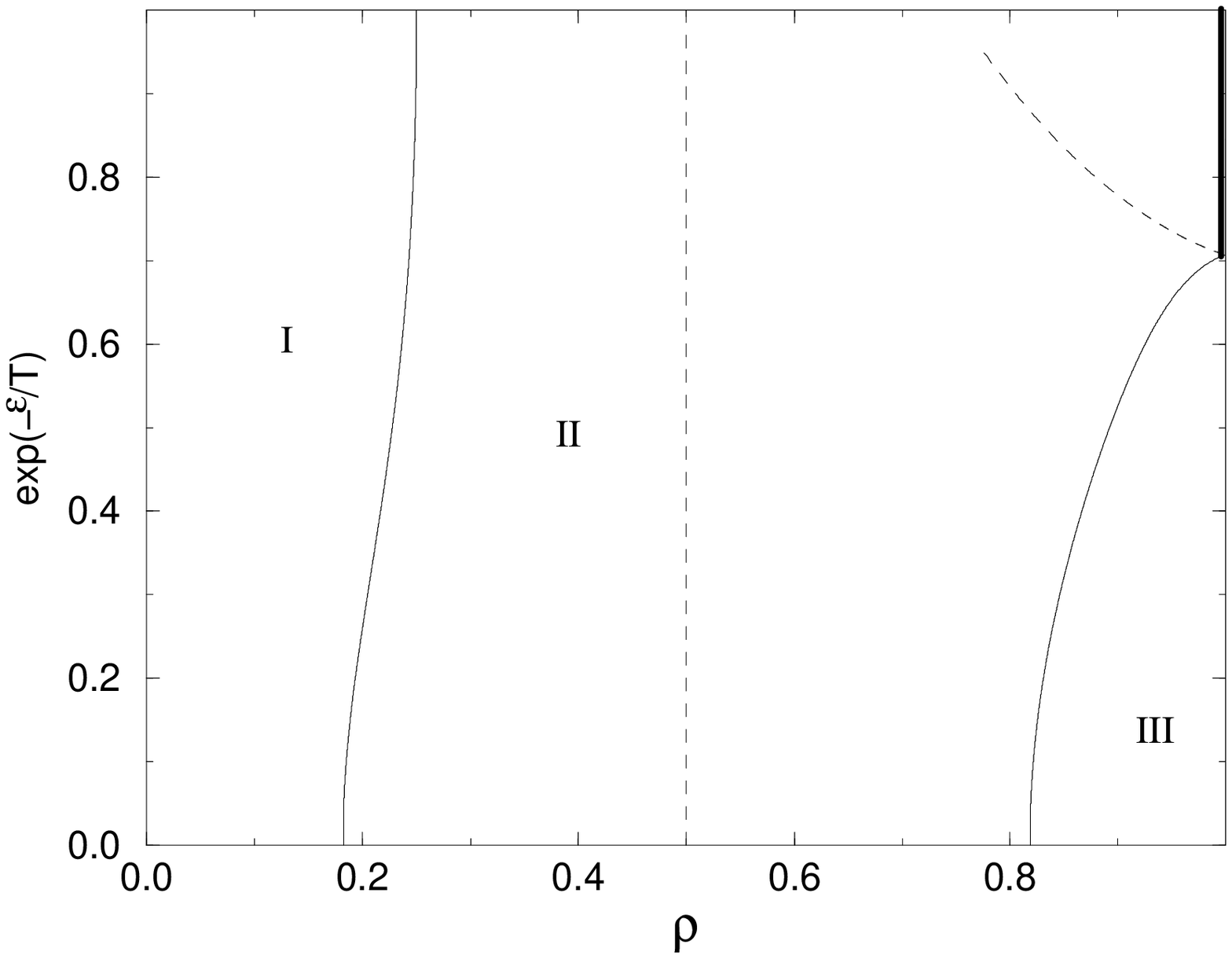,height=5in} }
\caption{Phase diagram of the loop gas in the $\rho-T$ plane; the solid
lines denote the phase boundaries; the thick line represents the
Lifshitz line; the dashed lines correspond to the disorder lines. The
three phases are defined in the text.}
\end{figure}


\begin{references}
\bibitem{Nienhuis}
B. Nienhuis in {\em Phase Transitions and Critical Phenomena},
Vol. {\bf 11}, ed. Domb and Lebowitz (Academic Press, New York 1987)

\bibitem{Baxter}
R. J. Baxter, {\em Exactly Solved Models in Statistical Physics}
(Academic Press, New York 1982)

\bibitem{deGennes}
P.-G. G. de Gennes, {\em Scaling Concepts in Polymer Physics} (Cornell
University Press, Ithaca 1979)

\bibitem{desCloizeaux}
J. des Cloizeaux and G. Jannink, {\em Polymers in Solution, their
Modelling and Structure} (Clarendon Press, Oxford 1990)

\bibitem{Wu}
E. M. Lieb and F. Y. Wu in {\em Phase Transitions and Critical Phenomena},
Vol. {\bf 1}, ed. Domb and Green (Academic Press, New York 1972)

\bibitem{Flory}
J. P. Flory, Proc. Roy. Soc. {\bf A 234} 60 (1956)

\bibitem{Pokrovski_Talapov}
V. L. Pokrovsky and A. L. Talapov, Phys. Rev. Lett. {\bf{42}}, 65-67 (1979)

\bibitem{wu}
F. Y. Wu, Phys. Rev. {\bf{168}} 539-543 (1968)

\bibitem{Blote}
H. W. J. Bl\"ote and H.J. Hilhorst, J. Phys. A: Math. Gen. {\bf{15}} 
L631-L637 (1982)

\bibitem{Stephenson}
J. Stephenson, Can. J. Phys. {\bf{47}} 2621 (1969) \\
J. Stephenson,  Can. J. Phys. {\bf{48}} 1724 (1970a)\\
J. Stephenson,  Can. J. Phys. {\bf{48}} 2118 (1970b) \\
J. Stephenson,  J. Math. Phys. {\bf{11}} 420 (1970c)

\bibitem{Houtappel}
R. M. F. Houtappel, Physica {\bf 16} 425 (1950)

\bibitem{Syozi}
I. Syozi in {\em Phase Transitions and Critical Phenomena},
Vol. {\bf 1}, ed. Domb and Green (Academic Press, New York 1972)


\bibitem{Dotsenko}
V. S. Dotsenko and V.S. Dotsenko, Advances in Physics {\bf{32}}, 129-172 (1983)

\bibitem{Samuel}
S. Samuel, J. Math. Phys. {\bf{21}} 2806-2814 (1980)

\bibitem{helfrich}
W. Helfrich, Z. Naturforsch. {\bf 33a}, 305 (1978).

\bibitem{garel}
T. Garel and J.M. Maillard, J. Phys. C {\bf{19}} L505-L511 (1986)

\bibitem{Saleur}
H. Saleur, J. Phys. A: Math. Gen. {\bf{19}} 2409-2423 (1986)


\end{references}
\end{document}